\documentstyle[11pt] {article}

\title{Switching of Geometric Phase in Degenerate Systems}
\author{A. Yahalom$^{a,b}$ and R. Englman $^{b,c}$\\
$^a$ Faculty of Engineering,\\
 Tel-Aviv University, Tel-Aviv 69978, Israel\\
$^b$ Research Institute,\\
 College of Judea and Samaria, Ariel 44284, Israel\\
$^c$ Department of Physics an Applied Mathematics,\\
Soreq NRC,Yavne 81810,
 Israel\\
e-mail: asya@post.tau.ac.il, englman@vms.huji.ac.il}

\begin{document}
\maketitle

\newcommand{\beq} {\begin{equation}}
\newcommand{\enq} {\end{equation}}
\newcommand{\ber} {\begin {eqnarray}}
\newcommand{\enr} {\end {eqnarray}}
\newcommand{\eq} {equation}
\newcommand{\eqs} {equations}
\newcommand {\er}[1] {equation (\ref{#1}) }

\begin {abstract}
The geometric and open path phases of a four-state system subject
 to time varying cyclic potentials are computed from the Schr\"{o}dinger
 equation. Fast oscillations are found in the non-adiabatic case. For parameter
values such that the system possesses degenerate levels, the geometric phase
 becomes anomalous, undergoing a sign switch.
A physical system to which the results apply is a
molecular dimer with two interacting electrons. Additionally, the sudden
switching of the geometric phase promises to be an efficient control
 in two-qubit quantum computing.


\bigskip
keywords: Berry's phase; Non-adiabaticity; Quantum computation

\end {abstract}

\section {Introduction}
Following the discovery of the topological (or geometric) phase acquired by
a system  that evolves adiabatically around a closed path in a parameter
 space  \cite{Berry}-\cite{Aharonov},
many instances of the phenomenon have been found either theoretically or
 in experiments \cite{Shapere}-\cite{Zwanziger}. A widespread quantum
 mechanical application of the theory is when an electron in a molecule
slowly evolves subject to a field due to (say) the host ions, such that the field
 varies periodically in time. In typical cases the electronic state changes sign
 after a full period \cite {Berry}. It is entirely obvious that for a pair of
electrons, which are independent except that both are subject to fields
 with the same period (though the fields may have otherwise different characters)
 the total electronic wave function (being a product of two individual wave
functions) will return to its original value after a period. Similarly,
when the field on the second electron is a constant or has a period
which is $\frac{1}{2}$ or $\frac{1}{4}$ (and so on) times the period for
 the first field then, after one complete revolution, the total wave function will
again acquire a change of sign. It is of interest to consider next what happens
 to the geometrical phase if one introduces an interaction between the two
 electrons. One would expect that the geometrical phase would not
be altered by the interaction. This is indeed the case for the model studied
 in this work, with one notable exception: When the two-electron states
 become degenerate by some coincidence
of the parameters in the Hamiltonian, then under suitable conditions
 the expected sign change becomes reversed. This phenomenon is studied
 in this work. Other aspects of states that have  degeneracies on or off the
 trajectories have been noted earlier in \cite {Wilzcek,Heiss}.
The topological phase change after a full revolution has been interpreted
in several previous publication as a surface integral in the parameter space
\cite{Berry,Shapere}. We find the acquired phase by a different method:
namely, by numerical integration of the time dependent Schr\"{o}dinger equation,
which is the appropriate method for circumstances that the surface integral is
not readily available. Moreover, this method, which was previously used for a
 single electron \cite {Eng1,Eng2,Eng3}, gives the entire open path phase,
 rather then the geometrical phase alone. Theoretical expressions and
 interpretations for the open path phase were given in \cite{Jain,
 Pati1}.
 The last section contains algebraic and numerically less convenient
  calculations.

\section {The physical model}
Brief descriptions of the model have been given previously
\cite{Baer1,Baer2}.
In essence, we consider a dimer consisting of two vibrating,
planar and possibly
dissimilar molecules lying upon each other.
A simple instance is two triatomic
molecules $ A_3$ - $B_3$. In addition to several tightly bound electrons,
each molecule contains one electron that is in an orbital doublet
 state, degenerate in the symmetric (equilateral triangle)
 configuration of its
 host. The molecules are sufficiently remote for the orbital doublets to be
 localized on the host molecule and also for overlap (and exchange) effects
 to be negligible. There remain vibronic interactions  between the electronic and the in-plane
vibrational motions, as well as an inter-electronic coupling. The former
(Jahn-Teller) interaction \cite {Eng4} is conveniently treated by the
formalism of Longuet-Higgins {\it et al} \cite {LH}, which represents the two
 electronic states by $e^{\pm i \theta_m}$ , where $\theta_m$
 an angular variable
and $m=1,2$ for the
 two molecules. In this work we shall modify this formalism by using as the
basis the direct product of two {\it real} representations, namely
 $\cos\theta_1 \cos \theta_2$, etc.
 The non-totally symmetric vibrational mode coordinates of the molecules are
 denoted by  $q_m$ $\cos \phi_m$ and $q_m$ $\sin \phi_m$ \cite {Eng4}. In terms
of the electronic and vibrational coordinates, as given above,
and the vibronic coupling strengths $G_m$,  the interaction may be simply
 expressed as
\beq
H_m = G_m \cos(2\theta_m -\phi_m ) \qquad m=1,2
\label {Hvibr}
\enq
>From elementary physical considerations, the vibronic constants  $G_m$ are
(at least approximately) proportional to the mode amplitudes $q_m$.
The interaction between electrons on the two molecular entities is familiar
from energy transfer and other subjects, and is expressed by terms that contain
jointly excitation and de-excitation operators on two molecules
 \cite {Dexter}. For simplicity we assume for the interaction term the large $U$
($t/U \rightarrow 0$),"anti-ferromagnetic Heisenberg" limit of the well-known
Hubbard Hamiltonian \cite {Hirsch} , having the product form

\beq
H_{12} = 2 G \cos(2\theta_1 -\phi_1 )\cos(2\theta_2 -\phi_2 )
\label {H12}
\enq
This is symmetric and even in the relative angular variables and is invariant
under a full rotation of either mode angular variable $\phi_m$. Analogously
 to Hubbard's $U$ which gives the
 exchange splitting between like and unlike spins on the same site, $G$ represents
 the electrostatic splitting between like and unlike vibronic states on the two
 molecules. $ G$  depends foremost on the distance between the molecules and only weakly on the
 molecular displacement coordinates.  In previous works \cite {Baer1,Baer2} the
 cases $G_m$ larger and smaller than $G$ were studied separately.(To exhibit
formally the weakness of overlap effects, one would introduce, both in the
wave functions and in the couplings, two more factors that depend on two further
coordinates, (say) $r_1$ and $r_2$, such that the overlaps between the $r_1$ and $r_2$
factors are negligible.)

The total Hamiltonian $H$ is the sum of \eqs  (\ref{Hvibr})
and (\ref{H12}).
\beq
H = H_1 + H_2 + H_{12}
\label {H}
\enq

We next represent the vibrations of the molecules as rotational motions
in the $ (q,\phi)$ -planes. (The potential surfaces in Figure 3.3 of \cite
{Eng4} admit of such motion.) Classically, this is equivalent to taking
$q_1 =constant$, $q_2 = constant$, $\phi_1 = \omega_1 t$,
$\phi_2 = \omega_2 t$,
where $t$ is time and $\omega_1$ and $\omega_2$ are the angular frequencies
 of the (small) molecular displacements.
The {\eq} that forms the basis of this work is the time dependent Schr\"{o}dinger
equation, with $\hbar=1$,
\beq
i \frac{\partial \Psi}{\partial t} = H \Psi
\label{SE}
\enq
 The two-electronic product states are labeled as follows:
\ber
\psi_1 &=& N \cos \theta_1 \cos \theta_2 \nonumber \\
\psi_2 &=& N \sin \theta_1 \cos \theta_2 \nonumber \\
\psi_3 &=& N \cos \theta_1 \sin \theta_2 \nonumber \\
\psi_4 &=& N \sin \theta_1 \sin \theta_2
\label {basis}
\enr
where the normalization constant, $N={1 \over \pi}$. In this basis the
 Hamiltonian finds the following representation:
\begin{eqnarray*}
H = \frac{1}{2} \times & & \qquad \qquad \qquad \qquad \qquad \qquad \qquad
\qquad \qquad \qquad \qquad \qquad \qquad \qquad
\end{eqnarray*}
\beq
{\tiny \left(\begin{array}{cccc}

G_1 \cos\phi_1 +G_2 \cos\phi_2 &
G_1 \sin\phi_1+G \sin\phi_1 \cos \phi_2 &
G_2 \sin\phi_2+G \cos\phi_1 \sin\phi_2 &
G \sin\phi_1 \sin\phi_2 \\

+ G \cos\phi_1 \cos\phi_2 & & & \\

G_1 \sin\phi_1+G \sin\phi_1 \cos\phi_2 &
-G_1 \cos\phi_1 +G_2 \cos\phi_2 &
G \sin \phi_1 \sin \phi_2 &
G_2 \sin\phi_2-G \cos\phi_1 \sin \phi_2 \\

& - G \cos \phi_1 \cos \phi_2 & & \\

G_2 \sin\phi_2+G \cos \phi_1 \sin \phi_2 &
G \sin\phi_1 \sin \phi_2 &
G_1 \cos \phi_1 -G_2 \cos \phi_2 &
G_1 \sin \phi_1-G \sin \phi_1 \cos \phi_2 \\

& & - G \cos \phi_1 \cos \phi_2 & \\

G \sin \phi_1 \sin \phi_2 &
G_2 \sin\phi_2-G \cos\phi_1 \sin \phi_2 &
G_1 \sin \phi_1-G \sin \phi_1 \cos \phi_2 &
-G_1 \cos \phi_1 -G_2 \cos \phi_2 \\

& & & + G \cos \phi_1 \cos \phi_2 \\
\end{array}\right)}
\label{Hmatr}
\enq
The eigen energies of this matrix, denoted by $\kappa_r^{AD} (r=1,..,4)$ and
 the corresponding eigen functions (the so-called adiabatic eigen states) are:
\ber
\kappa_1^{AD} &= \frac{1}{2}(G_1 +G_2 +G)
&\rightarrow \psi^{AD}_1 = {1\over \pi}
\cos(\theta_1 - {\phi_1\over 2})\cos(\theta_2 - {\phi_2 \over 2})
\nonumber \\
\kappa_2^{AD} &= \frac{1}{2}(-G_1 +G_2 -G)
&\rightarrow \psi^{AD}_2 = {1 \over \pi}
\sin(\theta_1 -{\phi_1\over 2})\cos(\theta_2 - {\phi_2 \over 2})
\nonumber \\
\kappa_3^{AD} &= \frac{1}{2}(G_1 -G_2 -G)
&\rightarrow \psi^{AD}_3 = {1\over \pi}
\cos(\theta_1 - {\phi_1 \over 2})\sin(\theta_2 - {\phi_2\over 2})
\nonumber \\
\kappa_4^{AD} &= \frac{1}{2}(-G_1 -G_2 +G)
&\rightarrow \psi^{AD}_4 = {1\over \pi}
\sin(\theta_1 -{\phi_1\over 2})\sin(\theta_2 - {\phi_2 \over 2})
\label{solution}
\enr

\section {Time evolution}

Recalling that $\phi_1 = \omega_1 t$ and $\phi_2 = \omega_2 t$, we substitute
the time dependent Hamiltonian matrix, \er {Hmatr}, into \er {SE} to obtain
 the solution $\Psi$ (which is now a four column-vector) by forward
 integration, using a "Mathematica" numerical algorithm. With the choice of
 $\omega_1 =1$ and $\omega_2$ a simple multiple of $\omega_1$, the periodicity
 is $2 \pi$ and the solutions exhibited in the following figures stretch over
a single period  of the Hamiltonian, $t=0 - 2\pi$. Beyond this, the solutions
can be obtained by symmetry. The initial value of
 $\Psi$ at $ t=0$ was taken as $(1,0,0,0)$. This choice implies no loss
 of generality, since any other solution can be derived from this through
a change of the parameters, shift of the time scale and superposition of solutions.
The analytic treatment of the Schr\"{o}dinger  equation including  the
  interaction, given in the last section ,
involves the solution of a quartic equation  which is not decomposable into two
quadratic equations. This stands in contrast to the one-electron two-
  state problem,
whose Schr\"{o}dinger equation involves the solution of a quadratic \cite {Rabi, Moore, Eng3}.
Thus the interaction  introduces in the problem a genuinely new element.

(A) Non-adiabatic effects.

 We first investigate the approach to the adiabatic limit, which is the condition
 under which the results of \cite{Berry} were derived.
 To achieve this limit, at least one of the coupling coefficients
 $G_1$, $G_2$ and $G$ has to be large. (Basically, their ratio to omega serves as
the measure of adiabaticity \cite{Eng2}. We define it by $A=|\frac{G_1}{\omega_1}|$)
 The approach to adiabaticity is
 exemplified in Figure 1, whose two curves have identical parameters
 except that all coupling strengths are reduced by a factor
 10 in the thinly drawn curve , thus not yet reaching the adiabatic limit.

\begin{figure}
\vspace{4cm}
\begin{picture}(1,1)
\end{picture}
\includegraphics{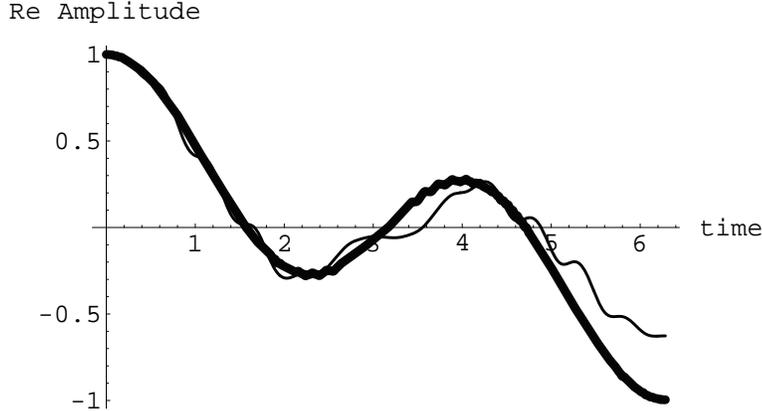}
\caption { Non-adiabaticity effects in the real part of the initially excited component,
 as a function of time. The frequencies on the two dimers are
$\omega_1=1 $ and $\omega_2=2$. The values of the coupling parameters are as follows.
Thick line: $G_1= -80$, $G_2=-160$, $G=40 $ (near adiabatic limit). Thin line:
$G_1=-8$, $G_2=-16$, $G=4$ (non-adiabatic case)}. The dynamic phase has been
subtracted in this and in the following figures.
\label {fig: nad1}
\end{figure}

 In either curve there is
  a sign change in the wave function at full revolution, but the lack of
 adiabaticity shows up in two ways:
 First,  in the state amplitudes it induces oscillations,
 whose periods depend on the coupling strength. These oscillations are the
ionic-field analogues of the Rabi oscillations (that are due to excitations by an
 electro-magnetic field). The former have been analyzed in detail in an earlier
paper on a two-state model \cite {Eng3}.
Secondly, the state amplitude does not reach -1, but only approximately -.75.
This reduction has already been quantitatively explained for the two state model in
 \cite{Eng1} and will not be further discussed here.
In the fully adiabatic limit, the solution can be read off from the
analytic expressions given in \er{solution}. Having started (by virtue of
the initial conditions imposed) with $\psi^{AD}_1$, we see that the thick curve
shown in Fig. 1 indeed approximates well to the expression
 $\cos(t/2) \cos(t)$, which is the amplitude of the first component in
$\psi^{AD}_1$, according to \er{basis} and \er{solution}.
Further solutions for the near-adiabatic limit are shown graphically
 in Figures 2 and 3.

\begin{figure}
\vspace{4cm}
\begin{picture}(1,1)
\end{picture}
\includegraphics{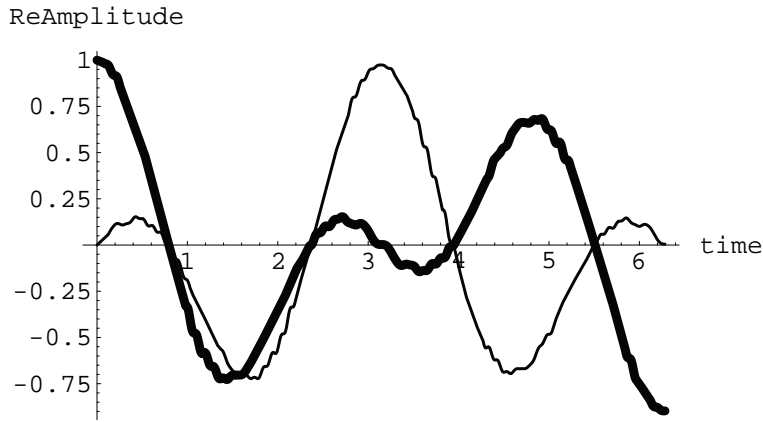}
\caption {Two-electron state amplitudes in a dimer.
The thick  line shows the time dependent amplitude of the first (initially excited component),
the thin line  that of the second component in Equation (5).
Frequencies: $ \omega_1=1 $, $\omega_2 =4$,
$G_1=-40$, $G_2=-80$, $G=16$ (near adiabatic limit)}
\label {fig: ad3}
\end{figure}

\begin{figure}
\vspace{4cm}
\begin{picture}(1,1)
\end{picture}
\includegraphics{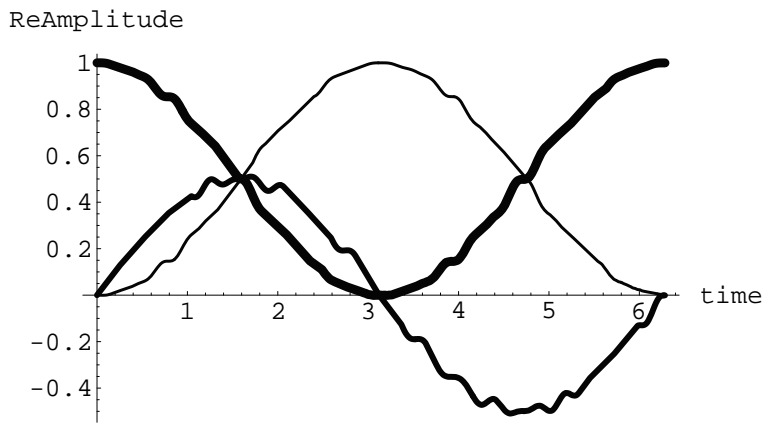}
\caption {Two-electron state amplitude in a dimer,
with both molecules subject to a periodic force.
After a full revolution the two electronic states each change their sign,
leaving the total state invariant. Frequencies: $\omega_1=1$, $\omega_2 =1$,
$G_1=-100$, $G_2=-200$, $G=40$ (near adiabatic limit).
Thick line: First, initially excited component.
Medium thick line: Second and third components.
Thin line: Fourth component.}
\label {fig: ad4}
\end{figure}

(B) Enters degeneracy.

Further computations in the adiabatic limit (large $G$'s) with other
 parameters likewise confirm the expressions in \er{solution}, as predicted
by the adiabatic theorem. As is well known, this theorem is based on the
non-crossing of energy levels \cite{Messiah}. In the absence of
 inter-electronic coupling, the two electrons move independently and, therefore,
even if there was a degeneracy in the system through the combination
of the levels, no effect would be felt. For a coupled two-electron system,
the degeneracy is expected to produce new effects.

This was indeed found. Thus, when in \er{H12} the inter-electron coupling was
 so adjusted that
\beq
\ G=-G_1
\label{degeneracy}
\enq
then the first two levels in \er{solution} become degenerate. The accompanying
 curves (a)-(c) in Figure 4 show the way the amplitude
 (of the first component) changes as
this degeneracy is traversed through subsequent sequence of changes in $G$.
 The results  are similar
when the changes are made in an electron-vibration coupling strength $G_1$.
The curves are computed in the near-adiabatic limit.

\begin{figure}
\vspace{4cm}
\begin{picture}(1,1)
\end{picture}
\includegraphics{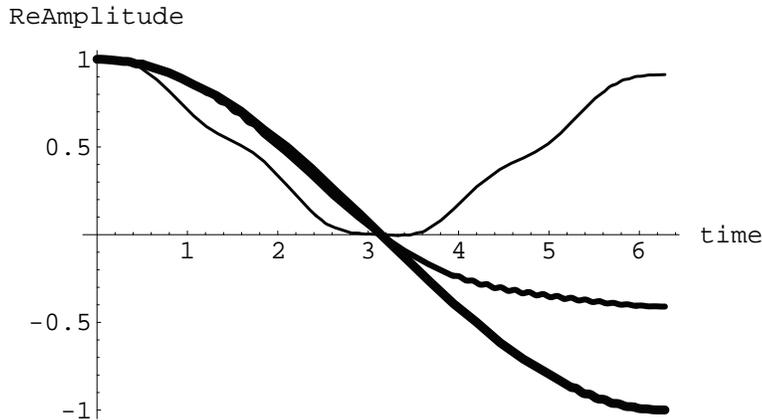}
\caption {Degeneracy and near-degeneracy effects in the first,
initially excited component.The following values are common to the three
    curves: $\omega_ 1=1,\omega_2=1,G_1=-40,G_2=-80$. The inter-electronic
 coupling $G$ is varied as follows: Thin-line: $G=36$ and $44$ (non-degenerate).
Medium thick line:$G=39.6$ and $40.4$ (near degenerate case).
Thick line: $G=40$ (exact degeneracy).}
\label {fig: ad5}
\end{figure}

One observes that with degenerate parameters, the amplitude keeps its sign at a
half period and achieves at full revolution a switched sign with respect to that
in the non-degenerate situation. The switch occurs abruptly as the coupling
strength(s) is (are) varied. We now turn to a discussion of this switch,
 regarding  its dependence on adiabaticity and its stability,
an application and the algebraic treatment of the
results.

\section {Discussion}

The preceding figure (fig. 4) was plotted in the near adiabatic limit when $A=
 |\frac{G_1}{\omega_1}|$ is large($=40$).In the extreme
adiabatic limit when $A =\infty$ only the symmetric ( thin )
line (ending at $+1$)  and the antisymmetric (thick) line
(ending at $-1$)  are possible,
the latter occurring at exact degeneracy and
the former for any values of the parameters that do not cause exact
degeneracy.When the adiabaticity parameter $A$ is finite, though large,
for large deviation from degeneracy the computed curve approaches the
symmetric curve and for small values of the deviation ,
the antisymmetric curve. It is possible to describe this effect
quantitatively, by finding a relation between the adiabaticity $A$ and that value  of the
deviation from precise degeneracy at which the real part of the
amplitude  at full revolution (at time $=2\pi$) is zero
(i.e., just half way between the values far from and exactly at
degeneracy). Attaching to  this value of the deviation from degeneracy the subscript $0$,
we find from our computation for the present model
\beq
|(G+G_1)_0|=0.54 \frac{|G_1|}{A}
\label{deviation}
\enq
(We recall that at $G+G_1=0$ one has a strict  degeneracy.) For a numerical
illustration of this result, when the adiabaticity $A = 40$
(this  being the absolute value of the ratio of the coupling
strength  $G_1 $ to the angular frequency of motion $\omega_1$,
as  e.g. when  $G_1=-40,\omega_1=1$),
then a deviation from degeneracy less than $0.54$  will cause in the
phase  a switch  whose magnitude is at least $\frac{\pi}{2}$ .In fig. 4,
the thin line is for a deviation $4$ that is larger than $0.54$ and the phase
switch is close to zero , the medium thick line is for a deviation of $.4$,
which is below the $0.54$ limit and the phase switch is larger than
$\frac{\pi}{2}$.Though \er{deviation} is derived from numerical fitting
(of over one and a half orders of magnitude in $A$ and to an average
accuracy of about five percent) for a specific model,
we expect that a qualitatively similar relation would hold in other cases
where the degeneracy causes a switch in the phase.
A deviation from (or removal of) the  degeneracy can also achieved by
adding off-diagonal terms ${\delta H_{12} }={\delta H_{21} }$ in
the $(1,2)$ and $(2,1)$ positions to he Hamiltonian matrix of
\eq (6).When these perturbation terms are constants their  effect  on
the switching is  given very closely by
\er{deviation}, when the deviation on the left hand side stands for $\delta H_{12}$,
or the  added matrix element.This represents a test for the stability of the
switch against a small deviation from the special form of the interaction
assumed in this work. (It would be tempting to combine the two findings into a single
formula, in which the left hand side of \er{deviation} is simply the splitting
of the two near degenerate levels, given by
\beq
splitting =((G+G_1)_0^2	+ 4 (\delta H_{12})^2)^{\frac{1}{2}}
\label {splitting}
\enq
but our numerical results indicate that the above formula holds without the
factor $4$.)

A further stability test was also made:
When  a constant was added to each of the terms in
top-right to bottom-left diagonal, the computed curves were not changed in a
 perceptible way, even when the constant reached $.2$ of the
(root mean square) value of the term.

A vibrating dimer, carrying one electron on either molecule, was used as a
 model system to introduce the formalism. We now turn to an alternative
application.

 The four  state formalism in this paper
 can be naturally interpreted as a two-qubit system. "Qubits" are two-state
 building-blocks for quantum computing. (An introductory account is found in
 \cite {DiVincenzo}.)The two-qubit situation is the minimal
 one to describe logical operations between control and target
\cite{Cirac,Monroe}, and it has also been shown to be
sufficient for that purpose \cite {Sleator}. For an identification of logical
operations in a two-qubit scheme, one can refer to \cite {Turchette}. The
correspondence between the time dependent description (or the Schr\"{o}dinger
 \eq  \ (\ref{SE}), on which the present work is based) and quantum computing
has been developed in \cite {Somaroo}. One of the several possible implementation
is the use of trapped ions in doublet quantum states \cite{King}.
as a control-no gate. This may be
 done as follows. We have seen  that as the first and second
components in \er{solution} are brought to
 degeneracy through the choice  $ G \rightarrow - G_1 $, then (upon full
 revolution) there is a sign-switch in the first component in \er {solution}.
Likewise, in the second (co-degenerate) component. There is, however, no
 switch in the third or in the fourth components (which are energetically
 distinct). Suitable combinations (e.g., sum and difference) of the first two
components can thus be used for representation of "Yes" and "No" operations,
with the choice of the coupling parameters acting as control.
 One important result of this paper is the abrupt
 switching of the geometric phase through a slight change in the coupling
 strength, as shown in Figure 4, and this can be conveniently used
 for manipulation on the target, or of the outcome, by external control. The
 abruptness of the switching is of some additional interest, in
 that the change in the total phase of the target, upon a  slight change of
 the parameter, is
more due to the geometric phase than to the change in the dynamic phase. This
means that the dynamic phase need not necessarily be subtracted from the total
 phase. Abrupt switching of the Pancharatnam phase for two polarized light
 beams was described and observed earlier \cite{Schmitzer}.

Further extension of our results might involve consideration of several
 (more than two) qubit states. This may be applicable
to quantum computation, in which
an important issue is how to achieve reduction of decoherence of phases by
random causes. It has thus been proposed that a possible way is the use of
superposition states composed of several qubits \cite{Shor, DiVincenzo}.

\section {Algebraic treatment}

  The solutions of the Schr\"{o}dinger \er{SE} and \er{Hmatr},
obtained in the text numerically,
can be exhibited through a sequence of(three)
 matrix multiplications
operating on a column (four) - vector $\vec \chi (t) $
which contains the initial
 conditions.  The vector is written in the representation of equation (6) as
\beq
\vec\chi (t) = \left (\begin{array} {c} e^{-i\kappa_1 t}\chi_1 (0)\\
 e^{-i\kappa_2 t}\chi_2 (0)\\ e^{-i\kappa_3 t}\chi_3 (0)\\
 e^{-i\kappa_4 t}\chi_4 (0) \end{array} \right)
\label {ksi}
\enq
and contains initial values needed to yield the initial values of
the solution and the eigen frequencies $\kappa_r$
 (of the Rabi-type) of a four matrix. In the extreme
adiabatic limit and excluding degeneracy. the eigen frequencies can be given
 explicitly and are
 identical with the eigenvalues  $\kappa_r^{AD}$ shown in \er{solution}.
 Non-adiabatic corrections to the (non-degenerate) eigenvalues are of the order
of $(\frac{\omega}{G})^2$ . The full solution is

\beq
\Psi = {\cal M} {\cal F} {\cal A} \vec \chi (t)
\label {product}
\enq

The matrix $M$ generates a complex representation when applied to the real
 basis shown in \er{basis}. Its form is

\beq
{\cal M} = \frac{1}{2} \left (\begin {array} {cccc} 1 & i & i & -1\\
1 & -i & i & 1 \\
1 & i & -i & 1 \\
1 & -i &-i & -1 \end{array} \right)
\label{M}
\enq
The next matrix ${\cal F}$ is diagonal and contains time exponents with the
 coefficient $\frac{1}{2}$, which is the source of the sign change upon a
 full revolution.

\beq
\cal F = \left(
\begin{array} {cccc}
e^{-{i\over 2} (\omega_1  + \omega_2) t} & 0 & 0 & 0 \\
 0 & e^{{i\over 2} (\omega_1  - \omega_2) t} & 0 & 0 \\
 0 & 0 &  e^{{i\over 2} (-\omega_1  + \omega_2) t} & 0 \\
 0 & 0 & 0 & e^{{i\over 2} (\omega_1  + \omega_2) t}
\end{array}
\right)
\label {F}
\enq

Lastly, ${\cal A}$ diagonalizes the following matrix with the eigenvalues
 $\kappa_r$ $(r=1,..4)$ appearing above.

\beq
H^{st} = \frac{1}{2} \left(
\begin{array} {cccc} -\omega_1 -\omega_2 & G_1 & G_2 & G\\
G_1 & \omega_1 -\omega_2 & G & G_2 \\
G_2 & G & -\omega_1 + \omega_2 & G_1\\
G & G_2 & G_1 & \omega_1 +\omega_2
\end {array}
\right)
\label {Hst}
\enq
To prove equation \er{product} substitute into the Schr\"{o}dinger
equation (4) and transpose the time derivative of the factor ${\cal F}$ in
\er{F} from the left to the right hand side.
In the adiabatic limit, when the frequencies are vanishingly small compared
to the coupling strengths,  $\cal A$ simplifies to
\beq
{\cal A}= \frac{1}{2} \left( \begin{array} {cccc}
1 & 1 & 1 & 1\\
1 &-1 & 1 &-1\\
1 & 1 &-1 &-1\\
1 &-1 &-1 & 1 \end{array}
\right)
\label{A}
\enq
When some of the adiabatic eigenvalues approach degeneracy, e.g.,
\beq
\kappa_2^{AD} \rightarrow \kappa_1^{AD}
\label{degeneracy2}
\enq
the situation changes dramatically. The corresponding non-adiabatic
eigenvalues differ from the adiabatic ones by a term that is linear in
 a frequency. Specifically, to a first order in the $\omega$'s,
\beq
\kappa_s -\kappa_s^{AD} =\pm \frac{1}{2} \omega_1 \qquad (s=1,2)
\label {kappa}
\enq
This half-frequency value in the $\kappa $ -exponents shown in \er{ksi}
 adds on a further factor of $-1$ upon a full revolution, thus accounting
for the sign switch found in the degenerate situation.
Also (as the numerical solutions show), two components in the four-vector
become vanishingly small throughout the time-range. With  the initial
 conditions chosen for our solution, the vanishing components are the
 second and the fourth. If we remove the corresponding rows and columns
from the Hamiltonian shown in \er {Hmatr}, the resulting two-by-two matrix
 becomes under the degeneracy condition $G_1=-G$,
\ber
H &=& \frac{1}{2} \left(\begin{array}{cc}
(G_2+G \cos\phi_1)\cos \phi_2 &
(G_2+G \cos\phi_1)\sin\phi_2  \\
(G_2+G \cos \phi_1) \sin \phi_2 &
-(G_2+G \cos\phi_1) \cos \phi_2 \end{array} \right)
\nonumber \\
& - & \frac{1}{2} G \cos \phi_1 \qquad {\rm (a \ scalar)}
\label {Hmatr1}
\enr
Since $(G_2+G \cos \phi_1)$ is a common factor for all elements
 in the matrix, it does not enter the component amplitudes and these contain only
the angle $\phi_2$. This represents a second algebraic justification for the
 phase switch upon degeneracy.

The relation in \er {degeneracy} represents a three dimensional surface of
 degeneracy in the four dimensional displacement-manifold \cite {Baer1, Baer2}.
 Actually, the surface
is (as expected) only two dimensional, for the following reason: The degeneracy
 involves states 1 and 2. An off-diagonal coupling between these states would
lift the degeneracy, even if this coupling would be so weak as not to affect
 the dynamic behavior. By our ignoring this coupling, we effectively put
it equal to zero, which represents a further relation between some of the
coordinates, say, between $q_1$ and $q_2$.In summary, we have investigated both the geometric phase (i.e., at
full revolution) and
 the interim open path phases for two interacting electrons, each
 in a doublet state, when they move in cyclically varying potentials. With
the choice of a product inter electron potential, the eigen energies are
constants along the motion. The dynamic phase (being the time integral
 of an eigen energy) can thus be conveniently subtracted off from the total
 phase. With more general interactions, one would need a more complicated
procedure for subtraction.
We have identified non-adiabatic effects (namely, high frequency oscillations
 and the incomplete return to the initial state after a full revolution).
 We then elucidated a switch of the sign of the wave-function as the system
develops a degeneracy in the two-electron energy levels. The switch was shown to
be quite sudden as the degeneracy situation was approached through varying
 the parameters in the Hamiltonian.

\section {Acknowledgment}

The authors thank Professor Alden Mead for critical readings of the manuscript and
Dr. Tal Mor for discussing some computing aspects.

\end{document}